\pdfoutput=1
\documentclass{natureprintstyle}

\usepackage{graphicx}
\usepackage{epsfig}
\usepackage{amssymb}
\usepackage{amsmath}
\usepackage{mathrsfs}
\usepackage{color}
\usepackage{url}
\usepackage{threeparttable}
\usepackage{multirow}
\usepackage{url}
\usepackage{pdflscape}

\newcounter{firstbib}
\newcommand{\lya}{Ly$\alpha$}
\newcommand{\tabincell}[2]{\begin{tabular}{@{}#1@{}}#2\end{tabular}}
\newcommand\figcaption{\def\@captype{figure}\caption}
\newcommand{\pc}{LAGER-$z$7OD1}

\newcommand\aap{Astron. Astrophys.}               
\newcommand\mnras{Mon. Not. R. Astron. Soc.}%
\newcommand\apjs{Astrophys. J. Suppl. Ser.}%
\newcommand\apj{Astrophys. J.}%
\newcommand\apjl{Astrophys. J.}%
\newcommand\pasp{Publ. Astron. Soc. Pacific}%

\newcommand\pasj{Publ. Astron. Soc. Jpn.}

\begin{document}

\title{A Lyman-$\alpha$ PROTOCLUSTER AT REDSHIFT 6.9}
\author{
  Weida Hu$^{1,2}$, 
  Junxian Wang$^{1,2}$, 
  Leopoldo Infante$^{3,6,9}$, 
  James E. Rhoads$^{4}$, 
  Zhen-Ya Zheng$^{5}$, 
  Huan Yang$^{3}$, 
  Sangeeta Malhotra$^{4}$, 
  L. Felipe Barrientos$^{6}$, 
  Chunyan Jiang$^{5}$,
  Jorge Gonz\'alez-L\'opez$^{3,9}$, 
  Gonzalo Prieto$^{6}$,
  Lucia A. Perez$^{7}$,  
  Pascale Hibon$^{8}$, 
  Gaspar Galaz$^{6}$,
  Alicia Coughlin$^{7}$,  
  Santosh Harish$^{7}$, 
  Xu Kong$^{1,2}$, 
  Wenyong Kang$^{1,2}$, 
  Ali Ahmad Khostovan$^{4}$, 
  John Pharo$^{7}$,
  Francisco Valdes$^{10}$, 
  Isak Wold$^{4}$, 
  Alistair R. Walker$^{11}$,  
  XianZhong Zheng$^{12}$}
\maketitle

\begin{affiliations}
  \item{CAS Key Laboratory for Research in Galaxies and Cosmology, Department of Astronomy, University of Science and Technology of China,
  Hefei, Anhui 230026, China; urverda@mail.ustc.edu.cn, jxw@ustc.edu.cn}
  \item{School of Astronomy and Space Science, University of Science and Technology of China, Hefei 230026, China}
  \item{Las Campanas Observatory, Carnegie Institution of Washington, Casilla 601, La Serena, Chile}
  \item{Astrophysics Science Division, Goddard Space Flight Center, 8800 Greenbelt Road, Greenbelt,  Maryland 20771, USA}
  \item{CAS Key Laboratory for Research in Galaxies and Cosmology, Shanghai Astronomical Observatory, Shanghai 200030, China}
  \item{Instituto de Astrof\'isica,  Facultad de F\'isica, Pontificia Universidad Cat\'olica de Chile, Santiago, Chile}
  \item{School of Earth and Space Exploration, Arizona State University, Tempe, AZ 85287, USA}
  \item{European Southern Observatory, Alonso de Cordova 3107, Casilla 19001, Santiago, Chile}
  \item{N\'ucleo de Astronom\'ia de la Facultad de Ingenier\'ia y Ciencias, Universidad Diego Portales, Av. Ej\'ercito Libertador 441, Santiago, Chile}
  \item{Community Science and Data Center/NSF’s NOIRLab, 950 N. Cherry Ave., Tucson, AZ 85719, USA}
  \item{Cerro Tololo Inter-American Observatory, NSF's NOIRLab, Casilla 603, La Serena, Chile}
  \item{Purple Mountain Observatory, Chinese Academy of Sciences, Nanjing 210023, China}

\end{affiliations}

\begin{abstract}
  Protoclusters, the progenitors of the most massive structures in the Universe, have been identified at redshifts up to 6.6 [refs\cite{Ouchi2005,Wang2005,Malhotra2005,Jiang2018,Harikane2019,Calvi2019}]. 
  Besides exploring early structure formation, searching for protoclusters at even higher redshifts is particularly useful to probe the reionization.
  Here we report the discovery of the protocluster \pc\ at redshift of 6.93, when the universe was only 770 million years old and could be experiencing rapid evolution of the neutral hydrogen fraction in the intergalactic medium\cite{Robertson2015, Kulkarni2019}.
  The protocluster is identified by an overdensity of 6 times the average galaxy density, and with 21 {narrowband} selected \lya\ galaxies, among which 16 have been spectroscopically confirmed. 
  At redshifts similar to or above this record, smaller protogroups with fewer members have been reported\cite{Castellano2018,Tilvi2020}.
  \pc\ shows an elongated shape and consists of two sub-protoclusters, which would have merged into one massive cluster with a present-day mass of $3.7\times10^{15}$ solar masses. 
  The total volume of the ionized bubbles generated by its member galaxies is found to be comparable to the volume of the protocluster itself, indicating that we are witnessing the merging of the individual bubbles and that the intergalactic medium within the protocluster is almost fully ionized. 
  \pc\ thus provides a unique natural laboratory to investigate the reionization process.
\end{abstract}

High redshift Lyman-$\alpha$ (\lya)-emitting galaxies (LAEs) are star-forming galaxies with strong \lya\ lines, which can be effectively selected with narrowband imaging surveys\cite{Zheng2017,Itoh2018,Hu2019}.
Aiming to build a statistical sample of LAEs at redshift $\sim$ 7, we are carrying out a deep narrowband imaging survey, Lyman Alpha Galaxies in the Epoch of Reionization (LAGER), utilizing the Dark Energy Camera (DECam, with a field of view of $\sim 3$ deg$^2$) on Cerro Tololo Inter-American Observatory (CTIO) Blanco 4m Telescope and a customized narrowband filter DECam-NB964. 
The central wavelength and full-width half-maximum of the filter are $\sim9642$ \AA\ and 92 \AA\ (see Fig. 1), corresponding to a redshift range of 6.89 -- 6.97 and a line-of-sight (LOS) scale of 26 cMpc. 
See Methods for more details.
In the LAGER COSMOS field, we obtained 47.25 hours narrowband exposure reaching a 5$\sigma$ detection limit of 25.2 magnitude {and a \lya\ sensitivity of $10^{42.65}$ erg s$^{-1}$}. Combining the deep narrowband image with the ultra deep broadband images from the Hyper Suprime-Cam Subaru Strategic Program (HSC SSP), we uniformly selected 49 $z\sim7$ LAEs\cite{Hu2019}. 
See Methods and papers\cite{Zheng2017,Hu2019} (hereafter Z17 and H19, respectively) for more details about the LAE selection. 

\begin{figure}[]
  \centering
  \includegraphics[width=3.3in]{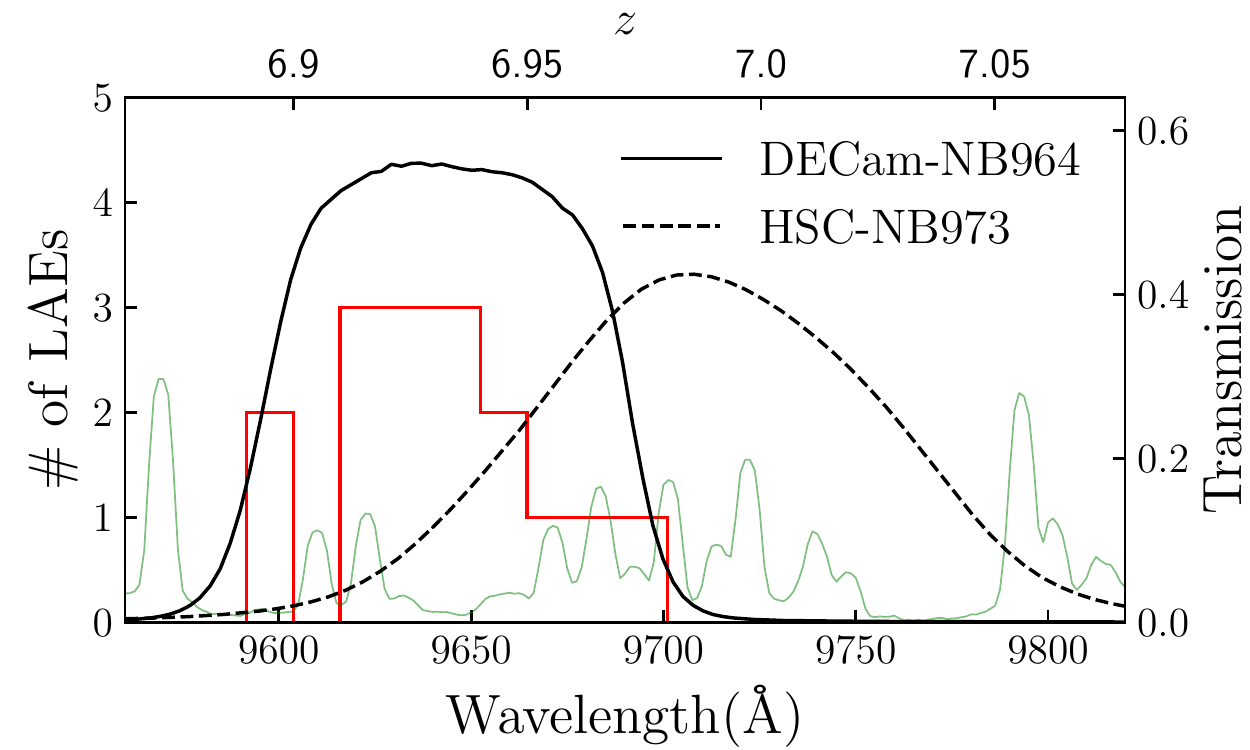}
  \caption{\textbf{Redshift distribution of spectroscopically confirmed LAEs in \pc.} {The red histogram shows the redshift distribution of 16 spectroscopically confirmed LAEs in \pc.} The black lines are the total transmission curve, including the full system response from atmosphere (at airmass of 1.2) to detector, of DECam-NB964 and HSC-NB973. Comparing with HSC-NB973, the transmission curve of DECam-NB964 is more like a boxcar with steeper wings. The sky OH emission lines are over-plotted in green.
  \label{fig1}}
\end{figure}

\begin{figure*}[]
  \centering
  \includegraphics[width=6.5in]{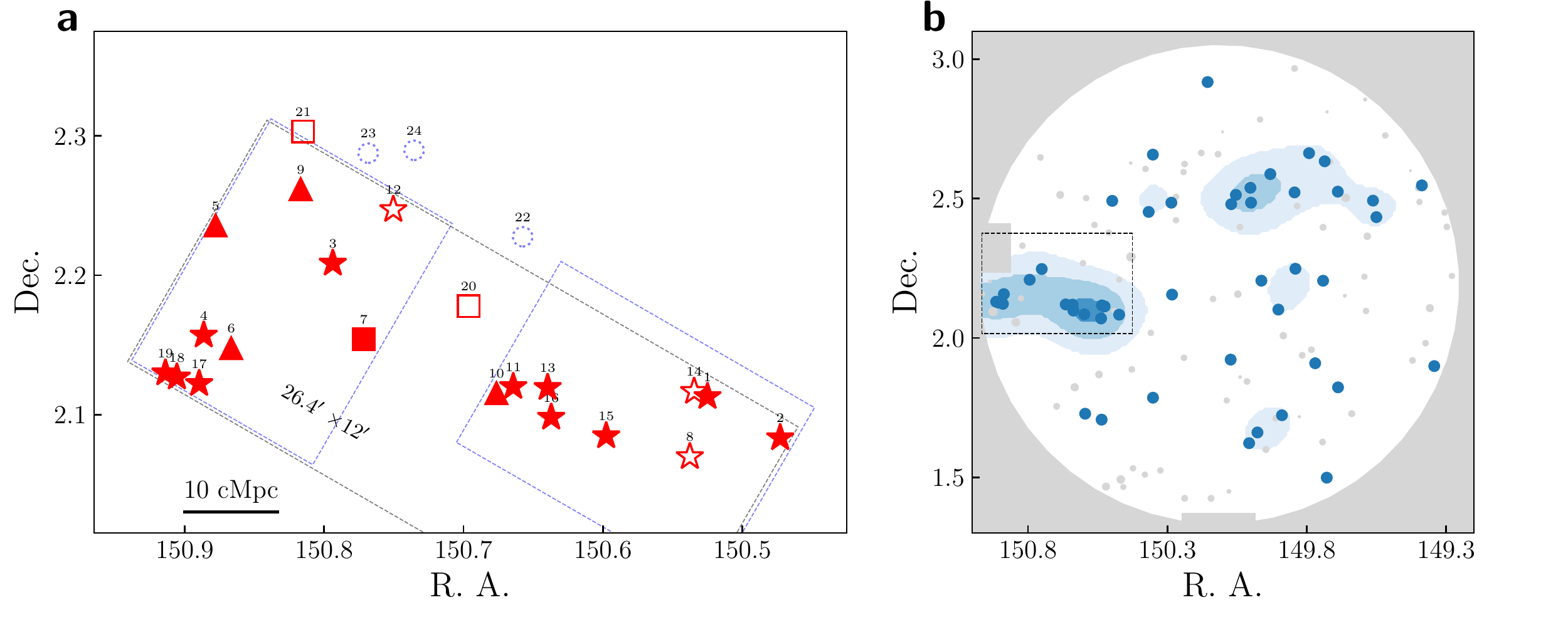}
  \caption{\textbf{Two-dimensional spatial distribution of LAEs in \pc\ at $z\sim7.0$.}  {\textbf{a,} The spatial distribution of member LAEs of \pc\ (a zoomed-in view of the black dashed rectangle region in \textbf{b}).
 Red symbols mark 21 member LAEs of \pc, including 14 candidate LAEs presented by H19 (red stars), three additional members selected with the aid of the HSC-NB973 image (red boxes), and four lower-grade photometric candidates that have been spectroscopically confirmed thus not presented in H19 (red triangles).
All members with spectroscopic confirmations are plotted with solid symbols while those no yet confirmed with open ones.}
Three candidate LAEs from HSC-NB973 [ref.\cite{Itoh2018}] in the area which are not detected in NB964 are marked as dashed blue circles. They are likely at higher redshifts (beyond the probe of DECam-NB964) and are not considered as members of \pc.
  The gray dashed rectangle ($26.4'\times12'$, corresponding to $\sim66\times30\ \mathrm{cMpc}^2$) represents the protocluster region. 
  Note that while all 21 red symbols (open and solid) are considered as members of \pc, only red stars (open and solid, those uniformly selected by H19) are utilized for over-density analyses. 
  The two blue dashed squares mark the two sub-protoclusters, each with a scale of $12'\times9'$.
  \textbf{b,} The spatial distribution of 49 LAGER $z\sim$ 7 LAEs (blue circles) in the whole COSMOS field presented in H19. The blue shadow contours show the local number densities of LAEs to highlight the overdense regions. Grey areas indicate the regions we masked out when we performed the analysis.
 \label{fig2}}
\end{figure*}

As narrowband imaging can constrain the redshift of LAEs to a very narrow range $\Delta z<$ 0.1, corresponding to a line-of-sight (LOS) distance of 30 -- 45 cMpc at $z = 6$ -- 8, it is also a promising approach to search for overdense structures, for example, protoclusters, in the early Universe\cite{Ouchi2005,Wang2005,Malhotra2005,Jiang2018,Harikane2019,Calvi2019}.
Fig. 2b shows the spatial distribution (blue circles) and number density (blue contours) of 49 LAGER $z\sim7$ LAEs in the whole COSMOS field as presented in H19. 
A high number density region (as marked by black dashed rectangle) is clearly revealed, containing 14 uniformly selected LAEs in H19 (see Suppl. Tab. 1 for the catalog). 
This overdense region (\pc) has a scale of $26.4' \times 12'$,  and a three-dimensional volume of $66 \times 30 \times 26$ cMpc$^3$. 
We calculate the galaxy overdensity of \pc\ following $\delta_g = (n-\bar{n})/\bar{n}$, where $n$ and $\bar{n}$ are the average number densities of LAEs in the \pc\ and the COSMOS field, respectively. 
We obtain the galaxy overdensity of \pc\ to be $\delta_g = 5.11^{+2.06}_{-1.70}$, which indicates \pc\ is a heavily overdense region, compared with the average galaxy number density.
See Methods for more details.
Up to now, candidate LAEs have been selected in four LAGER fields, that is COSMOS, CDFS, WIDE12, and GAMA15A (Z17, H19, Wold et al. in preparation). 
Among them, COSMOS is the unique one showing clear overdense region(s). 

The same field was also observed with another narrowband filter HSC-NB973\cite{Itoh2018}, the bandpass of which partially overlaps with that of DECam-NB964 (see Fig. 1). 
A hint of overdensity around \pc\ is also visible among the HSC-NB973 selected LAEs\cite{Itoh2018}, but not as strong as that seen in DECam-NB964. 
We stacked the DECam-NB964 image and HSC-NB973 image to improve the depth of the narrowband image and selected three more members of \pc.
We also plot in Fig. 2a four more members, which were selected as lower-grade candidates (compared with those presented in H19) but were later spectroscopically confirmed (see next paragraph). 
Note these 7 additional member galaxies are only used to illustrate the spatial profile of \pc\ (but not to calculate the overdensity), as they were not selected in a uniform and unbiased approach. See Methods for details.

Spectroscopic observations have been conducted to confirm the member LAEs, measure their redshifts, and remove potential contaminants which may show continuum or emission lines at blueward of 9600 \AA.
Three members have been confirmed in a previous study\cite{Hu2017}.
We carried out new spectroscopic followups using the Inamori Magellan Areal Camera and Spectrograph (IMACS) at the 6.5m Magellan I Baade Telescope (Feb. 6-8, 2017 and Feb. 21-23, 2018), and the Low Dispersion Survey Spectrograph 3 (LDSS3) at the 6.5m Magellan II Clay Telescope (January 10-11 and December 29-31, 2019). 
The average seeing during the observations was $\sim0.8''$. We carefully reduced the observed data and ruled out foreground identifications for member galaxies. 
Details of the data reduction are presented in Methods and a dedicated spectroscopic paper {in preparation} (along with identifications of LAEs outside of \pc\ and in other fields). 
In total, we have obtained spectroscopic confirmations for 16 member LAEs (red solid symbols in Fig. 2a). \lya\ lines were not detected in three additional members which were put on masks, however we are unable to rule them out as their \lya\ lines might incidentally overlap with sky lines, or their \lya\ line width be too broad to be detected\cite{Hu2017} ($>$ 500 km s$^{-1}$).  
The two- and one-dimensional spectra of the confirmed LAEs are presented in the Suppl. Fig. 1 and the redshift distribution in Fig. 1.

The scale ($66 \times 30 \times 26$ cMpc$^3$), overdensity ($\delta_g = 5.11^{+2.06}_{-1.70}$), and LOS velocity dispersion of spectroscopic confirmations ($\sim 765$ km s$^{-1}$) of our protocluster \pc\ are similar to those of the previously detected protoclusters at redshift of 5.7 -- 6.6 [refs\cite{Jiang2018,Harikane2019}] {and simulation predictions\cite{Overzier2009,Chiang2013}}. 
We estimate the total present-day cluster mass $M_{z=0}$ of \pc\ following the widely used formula\cite{Chiang2013}: $M_{z=0}=(1+\delta_m)\bar{\rho}V$, where $V$ is the volume of the protocluster, $\bar{\rho}$ ($3.88 \times 10^{10}\ M_\odot$ cMpc$^{-3}$) is the mean matter density of the universe, and $\delta_m$ the mass overdensity. 
$\delta_m$ is related to the observed galaxy overdensity through: $1 + b\delta_m = C (1 + \delta_g)$, where $b$ is the bias parameter and $C$ the correction factor for the redshift space distortion. 
For $\delta_g = 5.11$ at $z\sim7$, we find $C = 0.79$ and $\delta_m = 0.87$. 
The present-day mass $M_{z=0}$ of \pc\ is estimated to be $3.70^{+0.58}_{-0.51} \times 10^{15}\ M_\odot$, comparable to the mass of nearby COMA cluster\cite{Merritt1987} ($\sim 2 \times 10^{15}\ M_\odot$).
See Methods for details.

The 3D distribution of the spectroscopic confirmations is shown in Fig. 3. 
\pc\ shows an elongated shape and consists of two sub-protoclusters.
The overdensities of two sub-protoclusters are $6.76^{+3.77}_{-3.02}$ (left) and $9.34^{+4.21}_{-3.53}$ (right), respectively, where the boundaries of the two substructures are defined as the blue squares in Fig 2. 
If we treat the two substructures as isolated, their present-day masses are expected to be $1.39^{+0.32}_{-0.28}\times10^{15}\ M_{\odot}$ and $1.60^{+0.32}_{-0.31}\times10^{15}\ M_{\odot}$, respectively.

We further explore whether the protocluster \pc\ would collapse into a single cluster.
{Similar to previous works\cite{Cai2017,Chanchaiworawit2019}}, we estimate the linear overdensity of \pc\ to be $\delta_L = 0.54$ at $z\sim 7$ (Equation 18 of ref.\cite{Mo1996}). 
As the growth of linear perturbation $\delta_L$ is proportional to $t^{2/3}$, $\delta_L$ will be larger than the threshold $\delta_L > 1.69$ at $z = 2$, where $\delta_L = 1.69$ is the critical value of linear overdensity of a spherical perturbation at the time it collapses\cite{Jenkins2001}. 
Thus, we expect \pc\ collapses into a cluster at lower redshift. 
The discovery of \pc\ indicates that the formation of such large-scale structure had already begun by redshift 7.0, making it an ideal laboratory for understanding galaxy formation and large-scale structure formation.

\begin{figure}[]
  \centering
  \includegraphics[width=3.3in]{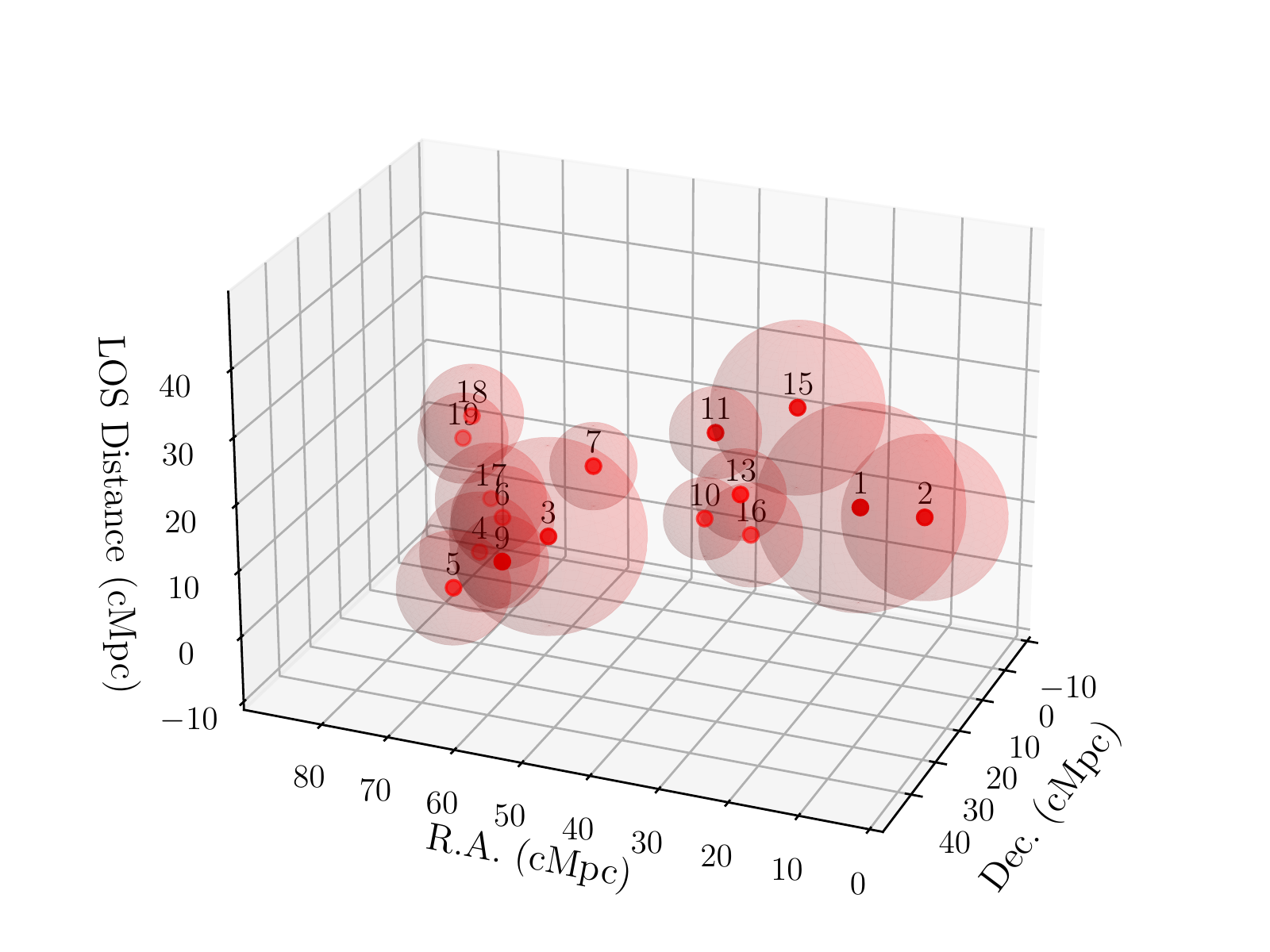}
  \caption{\textbf{3D spatial distribution of 16 spectroscopically confirmed LAEs in \pc\ at $z\sim7.0$.} 
  {The red points represent the 16 spectroscopically confirmed LAEs in \pc, which are shown as the solid symbols in Fig. 2a.}
  The translucent spheres denote the predicted ionized bubbles of LAEs. An interactive version can be found on https://www.lagersurvey.org/lager-z7od1.
  \label{fig3}}
\end{figure}

During the EoR, the hard UV photons that escaped from a galaxy could ionize the IGM and generate a HII region. 
The HII regions could gradually grow and merge with adjacent ones into sufficiently large ionized bubbles\cite{Tilvi2020,Castellano2018,Gnedin2000,RodriguezEspinosa2020}, which can reduce resonant scattering of \lya\ photons in the neutral IGM\cite{Malhotra2006,Dijkstra2007b}. 
The protoclusters in the EoR may lead the production of such bubbles because of their high number density of galaxies. 
On the basis of the relation between the bubble size and \lya\ luminosity in a simulation of reionization work\cite{Yajima2018}, we show the predicted bubbles in Fig. 3 and the bubble sizes in Suppl. Tab. 1. 
The summed volume of the ionized bubbles of all 21 LAEs is $6.58 \times 10^5$ cMpc$^3$, with the 4 most luminous ones (with $L_{Ly\alpha} > 2\times 10^{43}$ erg s$^{-1}$, i.e., LAE 1,2,3,15) contributing $60.3\%$ of the total ionized volume. 
This total ionized volume is even slightly larger than the volume of \pc\ ($5.15 \times 10^5$ cMpc$^3$). 
This demonstrates significant overlaps between individual bubbles, indicating the individual bubbles are in the act of merging into one or two giant bubbles (see Fig. 3). 
As a comparison, the total predicted volume of all the 49 uniformly selected LAEs in COSMOS field is $12.71 \times 10^5$ cMpc$^3$, corresponding to $11.1\%$ of the total volume surveyed by DECam-NB964. See Methods for details.
Such predicted giant bubbles are large enough to be resolved by future 21-cm programs, e.g., SKA1-Low with resolution of $\sim7.3$ arcmin at $z\sim 7.3$ [ref.\cite{Wyithe2015}], corresponding to $\sim 19$ cMpc. 

The merged bubble (with predicted size of $\gtrsim 30$ cMpc) could significantly increase the IGM transmission, and thus enhance the \lya\ visibility of member LAEs\cite{Dijkstra2007b}. 
Note {Z17 and H19} have revealed a bright-end excess in the \lya\ luminosity function in COSMOS field, also suggesting the existence of big ionized bubbles at $z \sim 7$ that reduce the opacity of neutral IGM around the luminous LAEs. 
Meanwhile, if the \lya\ transmission through the IGM has been significantly boosted in most LAEs in \pc, it may lead to larger \lya\ equivalent widths (EWs) of LAEs in the protocluster.
However, the expected larger \lya\ EWs is not seen, compared with the field LAEs in COSMOS  {(see Methods and Extended Data Fig. 1 for details)}, though the large uncertainties in the EW measurements and the small sample size prevent us from reaching a robust conclusion. 
One possibility is that high-redshift protoclusters are highly biased regions and might contain LAEs with physical properties deviating substantially from the field LAEs\cite{Harikane2019,Yajima2015}. 
It is yet unclear if the intrinsic \lya\ escape (prior to IGM scattering) in clustered LAEs is the same as that in field LAEs.

Moreover, the expected excess of close companions due to potentially enhanced \lya\ transmission, in or behind the large bubbles of the luminous LAEs ($L_{Ly\alpha} > 2\times10^{43}$ erg s$^{-1}$, i.e., LAE 1,2,3,15), is not seen (see Fig. 2a and 3). 
This is likely in part due to the possibility that while the biased dark matter halos can increase the galaxy merger/interaction, and thus enhance the star formation in the overdense region\cite{Harikane2019,Lee2017}, the feedback from the UV background may suppress the star formation in the nearby fainter galaxies\cite{Maio2016}. 

The discovery of the protocluster \pc\ provides an excellent opportunity to probe the rise and merging of ionized bubbles around the midpoint of EoR. 
Future deep and multi-band (HST, JWST, ALMA, etc) observations could reveal the detailed reionization processes, e.g., through searching for undetected \lya\ fainter galaxies partially responsible for the ionization budget, better constraining the \lya\ line EWs and the \lya\ profiles, measuring the \lya\ velocity offsets relative to their system redshifts, and mapping the star formations histories of the galaxies.

\begin{addendum}
 \item[Correspondence and request for materials]should be addressed to
   W. Hu and J. Wang (e-mail: urverda@mail.ustc.edu.cn, jxw@ustc.edu.cn).
 \item[Acknowledgements] 
 We appreciate the anonymous referees for the valuable comments and Zhen-Yi Cai, Zheng Cai, and Linhua Jiang for the informative discussions.
 The work is supported by National Science Foundation of China (grants No. 11421303 $\&$ 11890693 $\&$ 11773051 $\&$ 12022303), CAS Frontier Science Key Research Program (QYZDJ-SSW-SLH006), and CAS Pioneer Hundred Talents Program.
 US investigators on this work have been suppported by the US National Science Foundation through NSF grant AST-1518057; and NASA, through WFIRST Science Investigation Team contract NNG16PJ33C.
 
 This project used the data obtained with the Dark Energy Camera (DECam), which was constructed by the Dark Energy Survey (DES), 
 and public archival data from the Dark Energy Survey (DES). 
 
 Based on observations at Cerro Tololo Inter-American Observatory at NSF’s NOIRLab (NOIRLab Prop. ID: 2016A-0386, 2017B-0330; PI: Malhotra S.; CNTAC Prop. ID: 2016A-0610 ; PI: Infante L.), which is managed by the Association of Universities for Research in Astronomy (AURA) under a cooperative agreement with the National Science Foundation.
 
 This work includes data collected at the 6.5m Magellan Telescopes located at Las Campanas Observatory, Chile. We thank the scientists and telescope operators at Magellan telescope for their help.

 This paper makes use of software developed for the Large Synoptic Survey Telescope. We thank the LSST Project for making their code available as free software at  http://dm.lsst.org
 
 Based [in part] on data collected at the Subaru Telescope and retrieved from the HSC data archive system, which is operated by Subaru Telescope and Astronomy Data Center at National Astronomical Observatory of Japan.
 
 {\it Facilities:} {Magellan:Baade (IMACS), Magellan:Clay (LDSS3), Blanco (DECam), Subaru (HSC)}
 \item[Author contributions]
 W.H. and J.W designed the layout of this paper. W.H. reduced the data, performed scientific analysis and wrote the manuscript. 
J.W. co-led the scientific interpretation and manuscript writing.  L.I. led the observing proposals which yielded new spectroscopic identifications presented in this work.
W.H., L.I., H.Y., J. G. \& G.P. conducted these observations.
 All authors discussed the results and commented on the manuscript.

 \item[Competing Interests] The authors declare that they have no competing financial interests.
 
 \item[Data availability] The candidate selection is based on the following images in COSMOS field: DECam-NB964 {(NOIRLab Prop. ID: 2016A-0386, 2017B-0330; CNTAC Prop. ID: 2016A-0610)}, HSC SSP program, and HSC-NB973 {(Prop. ID: S16B-001I)}, which are available at \url{http://archive1.dm.noao.edu/}, \url{https://hsc-release.mtk.nao.ac.jp/doc/}, and \url{https://hsc-release.mtk.nao.ac.jp/doc/index.php/chorus/}, respectively. 
 The spectroscopic datasets and the datasets generated or analysed during this study are available from the corresponding author upon reasonable request.
 {The LAE catalog for \pc\ used in this study has been published in the Supplementary Information. }

 \item[Code Availability Statement] {Codes used in this study are not publicly released yet but are available from the corresponding author on reasonable request.}
\end{addendum}

\clearpage



\clearpage

\begin{methods}

  Throughout this study, we adopt the recent Planck cosmological parameters\cite{Planck2018VI}: $\Omega_m = 0.3111$, $\Omega_\Lambda = 0.6889$ and $H_0 = 67.66$ km s$^{-1}$ Mpc$^{-1}$, where $\Omega_m$ and $\Omega_\Lambda$ are the densities of total matter and dark energy and $H_0$ is the Hubble constant.\\

  \subsection*{Lyman Alpha Galaxies in the Epoch of Reionization (LAGER) survey: }
  {LAEs are promising probes for characterizing the cosmic reionization\cite{Malhotra2004,Furlanetto2006A,McQuinn2007,Ouchi2010,Hu2010,Zheng2017,Jiang2017,Chanchaiworawit2017,Konno2018,Itoh2018,Hu2019,Higuchi2019,Taylor2020,Jung2020}. 
  We are carrying out a large-area narrowband imaging survey, Lyman Alpha Galaxies in the Epoch of Reionization (LAGER), to search for the LAEs at $z\sim 7.0$, using the Dark Energy Camera (DECam) installed on the Cerro Tololo Inter-American Observatory (CTIO). 
  We designed and procured a narrowband filter (DECam-NB964) for the LAGER survey, with a central wavelength of $\sim9642$ \AA\ and FWHM of $\sim$ 92 \AA\ to avoid the atmospheric absorption and strong OH emission lines. 
  The filter DECam-NB964 was installed on the DECam system in December 2015.
  Owing to the large FoV ($\sim 3$ deg$^2$ ) and red-sensitive camera, LAGER is one of the most efficient surveys in searching for LAEs in the EoR. See the filter design paper\cite{Zheng2019} for more details.} 
  We adopt a "wedding cake" observing strategy with two deep fields aiming to discover faint LAEs and several shallower fields aiming to discover numerous luminous LAEs. 
  Up to now, candidate LAEs have been selected in 4 fields, including COSMOS, CDFS, WIDE12, and GAMA15A. \\\\
  
  \subsection*{\pc\ member galaxies: \\\\}
  {\bf a. Member galaxies from H19:}\\
  The 14 LAEs from H19 were selected by narrowband technique. 
  This technique is widely used in literature and has been proven effective at searching for LAEs. 
  Briefly, the selection criteria in H19 include:
  (1) the signal-to-noise ratio (S/N) of DECam-NB964 signal is larger than 5; 
  (2) DECam-NB964 excess over the underlying broadband to ensure the rest frame equivalent width (EW) of \lya\ is larger than 10 \AA; 
  (3) non-detection in bluer broadbands (we adopt the recently release HSC SSP ultradeep broadband images\cite{Aihara2018}). 
  We visually inspected each LAE candidate to remove possible foreground galaxies and spurious objects, such as satellite trails, cosmic rays, etc. 
  Finally, we obtained a clean sample of 49 LAEs in COSMOS field and a clear overdense region is revealed with 14 LAEs.\\\\
  {\bf b. Additional members:}\\
  The same field was also observed with another narrowband filter HSC-NB973\cite{Itoh2018}, the bandpass of which partially overlaps with that of DECam-NB964 (see Fig. 1). 
  A hint of over-density around \pc\ is also visible among the HSC-NB973 selected LAEs\cite{Itoh2018}, but not as strong as that seen in DECam-NB964. 
  Among the 8 HSC-NB973 selected LAEs in the area, four (LAE-1,2,11,15) were detected in DECam-NB964 and presented by H19. 
  An additional source (LAE-20) shows tentative signal ($S/N=4.7$) in DECam-NB964 image (see also next paragraph), while the remaining 3 (LAE-22,23,24; blue dashed circles in Fig. \ref{fig2}; namely {HSC-z7LAE24,6,16 respectively in ref.\cite{Itoh2018}}) are invisible in DECam-NB964. 
  {The latter three had not been spectroscopically observed, and are candidate LAEs likely at slightly higher redshifts beyond the probe of NB964 (see Fig. 1).  
  At the current stage, we do not consider these 3 LAEs as member galaxies of \pc\ as they may locate at slightly but sufficiently higher redshifts than that of the structure.}
  
  We further stack the DECam-NB964 and HSC-NB973 images to search for fainter LAEs located within the common volume sampled by two filters, and include three more candidates (LAE-7, LAE-20 and LAE-21). 
  We also plot in Fig. 2a four more DECam-NB964 LAEs (LAE-5, 6, 9, 10) which were selected as lower grade candidates (comparing with those presented in H19) but were later spectroscopically confirmed.\\\\
  
  \subsection*{{The overdensity in \pc:}} 
  We estimate the overdensity as defined by $\delta_g=(n-\bar{n})/\bar{n}$, where $n$ and $\bar{n}$ are the average LAE number densities in the \pc\ and the COSMOS field, respectively. 
  The number density of LAEs in the COSMOS is $\bar{n}\sim 0.0072^{+0.0012}_{-0.0010}$ arcmin$^{-2}$ (49 over 1.9 deg$^2$) and the number density of LAEs in the \pc\ is $n\sim0.0442^{+0.0152}_{-0.0116}$ arcmin$^{-2}$ (14 over 0.088 deg$^2$).
  The errors are calculated based on the Poisson errors of the LAE sample size. 
  The galaxy overdensity of \pc\ is thus $\delta_g = 5.11^{+2.06}_{-1.70}$. 
  The LAE sample suffers incompleteness during the detection and selection procedures (see H19 for details). 
  However, as the narrow- and broad-band images utilized for LAE detection and selection have rather uniform depths throughout the COSMOS field, the incompleteness is constant over the field, and thus, cancels out in the calculation of the overdensity.
  Note we use only the LAE sample in COSMOS field from H19 for the calculation of overdensity. The additional members to \pc\ aforementioned were excluded from such analyses as they were not uniform selected.  

  We adopt an enclosing rectangle as the boundary of the \pc\ (see the gray dashed rectangle in Fig. 2a) to calculate its volume and overdensity. 
However, the selection of the boundary is kind of arbitrary and may differ from the intrinsic shape of the protocluster. This could introduce systematic errors to the estimations of the volume thus overdensity and present-day mass. For instance, we do not see member LAEs in the lower-right region of the rectangle (see Fig. 2a) but this region contributes $\sim30\%$ of the volume.
If we exclude this void region, the overdensity would be increased to $\sim$ 7.73. Moreover, we simply calculate the overdensity using the light-of-sight scale (26 cMpc) probed by NB964, and the protocluster may have more members out of that range. Nevertheless, the effect of the boundary selection on the present-day mass estimation is moderate, as further discussed below.

  As aforementioned, the COSMOS field is unique among four LAGER fields, showing clear overdense region(s). The average LAE number density in the COSMOS field could be biased by cosmic variance, and such effect may also affect the calculation of the overdensity. 
  We compare the luminosity function (LF) of LAEs in the COSMOS field with those in other three LAGER fields and find the LFs to agree within $1\sigma$ Poisson errors (Wold et al. in preparation).
 We integrate the LFs in the luminosity range of $10^{42.65}-10^{43.65}$ erg s$^{-1}$ and find the derived average LAE densities from the four fields agree within $15\%$. 
 Thus the field-to-field variation has no significant effect on the calculation of overdensity. 
 
We finally note that we can not rule out the possibility that a small fraction of the uniformly selected 49 candidate from H19 are actually not real LAEs, but contaminants (such as noise spikes in narrowband image, variable sources, or foreground emission line galaxies). The total number of contaminated foreground emission line galaxies (H$\alpha$, [OIII] and [OII]) was estimated to be 0.82 in COSMOS, thanks to the ultradeep broadband images available\cite{Hu2019}. 
Considering the contaminants are unlikely spatially associated with the protocluster and should distribute randomly, excluding such contaminants (even if possible) from the calculation would yield even higher overdensity. 
\\\\

  \subsection*{Spectroscopic observation and data reduction:}
  The three brightest LAEs in \pc\ have been spectroscopically confirmed with IMACS on February 6-8, 2017 [ref.\cite{Hu2017}]. 
  We carried out spectroscopic followups for more LAEs in \pc\ using IMACS at the 6.5m Magellan I Baade Telescope (February 21-23, 2018), and LDSS3 at the 6.5m Magellan II Clay Telescope (January 10-11 and December 29-31, 2019). 
  For IMACS observations, we used the f/2 camera (with a FoV of $27'$ diameter) and the 300-line red-blazed grism. 
  For LDSS3, we used VPH-Red grism and OG590 filter to eliminate second order contamination. 
  Comparing with IMACS, LDSS3 has a smaller FoV ($8.4'$ diameter) but relatively higher efficiency at 9600 -- 9700 \AA. 
  Slitwidth of $1''$ was adopted for both instruments.
  The spectral reduction was performed using {\it{COSMOS3}}\cite{Dressler2011,Oemler2017} and the single-epoch spectra are average with weights selected to maximize the $S/N$ of the coadded spectra. 

  {The result 1D spectra (both IMACS and LDSS3) have spectral resolution of $\sim$ 6 \AA.
  We carefully examine the 2D spectra of all spectroscopic targets and identify 13 sources as LAEs based on single line detections. 
  Including the three previous confirmations\cite{Hu2017}, we now have spectroscopic confirmations for 16 member LAEs in \pc\ (red solid symbols in Fig. 2a). 
 The spectra of these 16 LAEs are presented in Suppl. Fig. 1. }
  
Single line detections (no other lines, no continuum) might still be contaminated by foreground strong emission line galaxies, e.g., H$\alpha$, [O$_{\mathrm{III}}$], and [O$_{\mathrm{II}}$]. 
Due to the limited spectral quality (and partial overlap with sky lines for some of them) we are unable to secure the characteristic asymmetric line profile\cite{Kashikawa2006} (with a red wing) of high-z \lya\ lines for many sources.
Meanwhile while some lines are too narrow to be [O$_{\mathrm{II}}$] doublet, for some broader ones [O$_{\mathrm{II}}$] can not be completely ruled out based on the line profile alone\cite{Hu2017}.
However, the contamination rate is expected to be low. For example, recent spectroscopic survey of high-redshift LAEs at $z\sim5.7$ reports a low contamination rate of $<10\%$ in their spetroscopic detections\cite{Ning2020}. 
Even we consider a contamination rate of 10\%, our single line identifications would be reliable for most sources.
More critically and fortunately, in COSMOS ultra deep broadband images are available to rule out almost all low-z interlopers of emission line galaxies, and we expect the sample of H19 include only $\sim$
0.14 ([O$_{\mathrm{II}}$]), 0.52 ([O$_{\mathrm{III}}$]), and 0.16 (H$\alpha$) low-z emission line galaxies over the whole COSMOS field\cite{Hu2019}.

  LAE-8, 12, and 14 were also spectroscopically observed, but not yet confirmed. 
  The non-detections of the \lya\ in their spectra do not necessarily rule them out, as their \lya\ lines might incidentally overlap with sky lines, or the velocity dispersions of their \lya\ lines could be too broad to be detected ($>$ 500 km s$^{-1}$). {Note we do not detect either any signals (lines, continuum) indicative of foreground sources in their spectra.
These candidates which are spatially associated with the protocluster are more likely real LAEs instead of contaminations (such as variable sources or noise spikes in the narrowband image). 
This is because the area of \pc\ is only $\sim$ 1/21 of the whole COSMOS filed, thus even if a small fraction of the 49 candidates selected over the whole field are indeed contaminants, we would expect no more than one of them within the area of \pc\ (assuming the contaminants randomly distribute over the field).
Therefore we opt to keep all three of them as valid candidates.
We further note that even if we were able to secure all such contaminations over the whole field, excluding such contaminations from the calculation would yield even higher overdensity.}
 \\\\

  \subsection*{Present-day mass of \pc: \label{subsss}}
  We estimate the total present-day cluster mass $M_{z=0}$ of \pc\ following the widely used formula\cite{Chiang2013,Steidel1998}: $M_{z=0}=(1+\delta_m)\bar \rho V$,  where $V$ is the volume of the protocluster, $\bar \rho\ (3.88 \times 10^{10}\ M_{\odot}$) is the mean matter density of the universe, and $\delta_m$ is the mass overdensity. 
  $\delta_m$ is related to the observed galaxy overdensity through: $1+b\delta_m = C(1+\delta_g)$, where $b$ is the bias parameter and $C$ the correction factor for the redshift space distortion, $C=1+f-f(1+\delta_m)^{1/3}$ and $f=\Omega_m z ^{4/7}$. 
  The bias parameter is assumed to be {$b=4.54\pm0.63$} (measured using $z\sim 6.6$ LAEs\cite{Ouchi2018}). For $\delta_g = 5.1$ at $z\sim 7$, we find $C = 0.79$ and $\delta_m = 0.87$. 
  Thus, the present-day mass $M_{z=0}$ of \pc\ is estimated to be $3.70^{+0.58}_{-0.51} \times 10^{15}\ M_\odot$ (where the errors are derived through simulating the fluctuations of the galaxy overdensity $\delta_g$ and bias parameter $b$). 
  {
  As aforementioned, the selection of the boundary is kind of arbitrary and could introduce as much as $30\%$ uncertainty to the overdensity estimation. However, this effect is moderate when estimating the present-day mass of the \pc. Decreasing the volume by $30\%$ would yield a $17\%$ lower $M_{z=0}$ and increasing the volume by $30\%$ would yield a $16\%$ higher $M_{z=0}$. 
  }
  The bias parameter might have been underestimated since we adopted the value at $z\sim 6.6$, and specifically, an increase of $b$ from 4.5 to 5.5 will result in a $\sim 7\%$ decrease of the estimated $M_{z=0}$. \\\\
  
  \subsection*{Bubble size estimation:}
  Previous studies\cite{Yajima2018} presented a semi-numerical simulation to investigate the relation between the bubble size and \lya\ luminosity of high-redshift LAEs in EoR. 
  In the simulation, the star formation rate (SFR) was assumed proportional to the growth rate of the dark matter halo, the escape fraction of ionizing photons was assumed to be 0.2, and the ionizing photon emissivity was calculated based on the star formation history of the galaxy using the population synthesis code {\it{STARBURST99}}\cite{Leitherer1999}. 
  Finally, the evolution of ionized bubble and \lya\ luminosity (derived from the ionizing photons which do not escape) were obtained after calculating the radiative transfer in the IGM. 
  Based on the relation between the bubble size and \lya\ luminosity (at $z=8$, Fig. 15 in ref.\cite{Yajima2018}), we show the predicted bubble (as translucent spheres) in Fig. 3 and the predicted bubble size in Suppl. Tab. 1 for the spectroscopically confirmed LAEs in \pc. 
 
  Note the derived bubble sizes are model dependent. Ref.\cite{Yajima2018} has assumed a constant mean IGM density outside an HII bubble with a clumping factor of $C$ = 3 considered to take account of the density fluctuation. 
  While increasing the clumpiness $C$ would not significantly decrease the bubble size\cite{Cen2000}, ref.\cite{Yajima2018} has pointed out that the higher IGM density near the virial radius may reduce the predicted bubble sizes. 
  The predicted bubble size is sensitive to the Lyman continuum escape fraction which was assumed to be a constant of 0.2 in ref.\cite{Yajima2018}.
  But note the observational results at $z<4$ suggest that only a small fraction of galaxies has a high escape fraction of $>10\%$\cite{Vanzella2010,Grazian2017,Bian2020}, 
  and the ionizing continuum escape fraction could be mass-dependent\cite{Finkelstein2019,Ma2020}.
  Moreover, in the model of ref.\cite{Yajima2018}, the \lya\ escape fraction was assumed to be a constant of 0.6, and both the bubble size and \lya\ luminosity tightly correlate with galaxy stellar mass. 
  However, it is known that high redshift LAEs on average are low mass and young galaxies, i.e., the \lya\ escape fraction is mass and stellar age dependent (e.g. refs\cite{Finkelstein2009,Lai2008,Cai2014}). 
  Consequently, our LAEs could be significantly less massive and younger than the model predictions of ref.\cite{Yajima2018}, and thus would be expected to have considerably smaller bubble sizes. 
 
  Nevertheless, considering the mean neutral hydrogen fraction at $z \sim 7.0$ ($x_{HI} =$ 0.2 -- 0.4, H19) and the significant overdensity of \pc, it is reasonable to believe that the IGM in \pc\ was close to fully ionized at $z \sim 7.0$. 
  However, it is yet uncertain whether the member LAEs we detected alone can produce such a giant bubble, as their predicted bubble sizes are remarkably model dependent. 
  {In the cases aforementioned, more undetected \lya\ fainter galaxies (with lower star formation rates and/or \lya\ escape fraction) could have contributed to the reionization around \pc.}
  \\\\

  \addtocounter{figure}{-3} 
\begin{figure}
  \renewcommand{\figurename}{Extended Data Figure}
  \centering
  \includegraphics[width=3.3in]{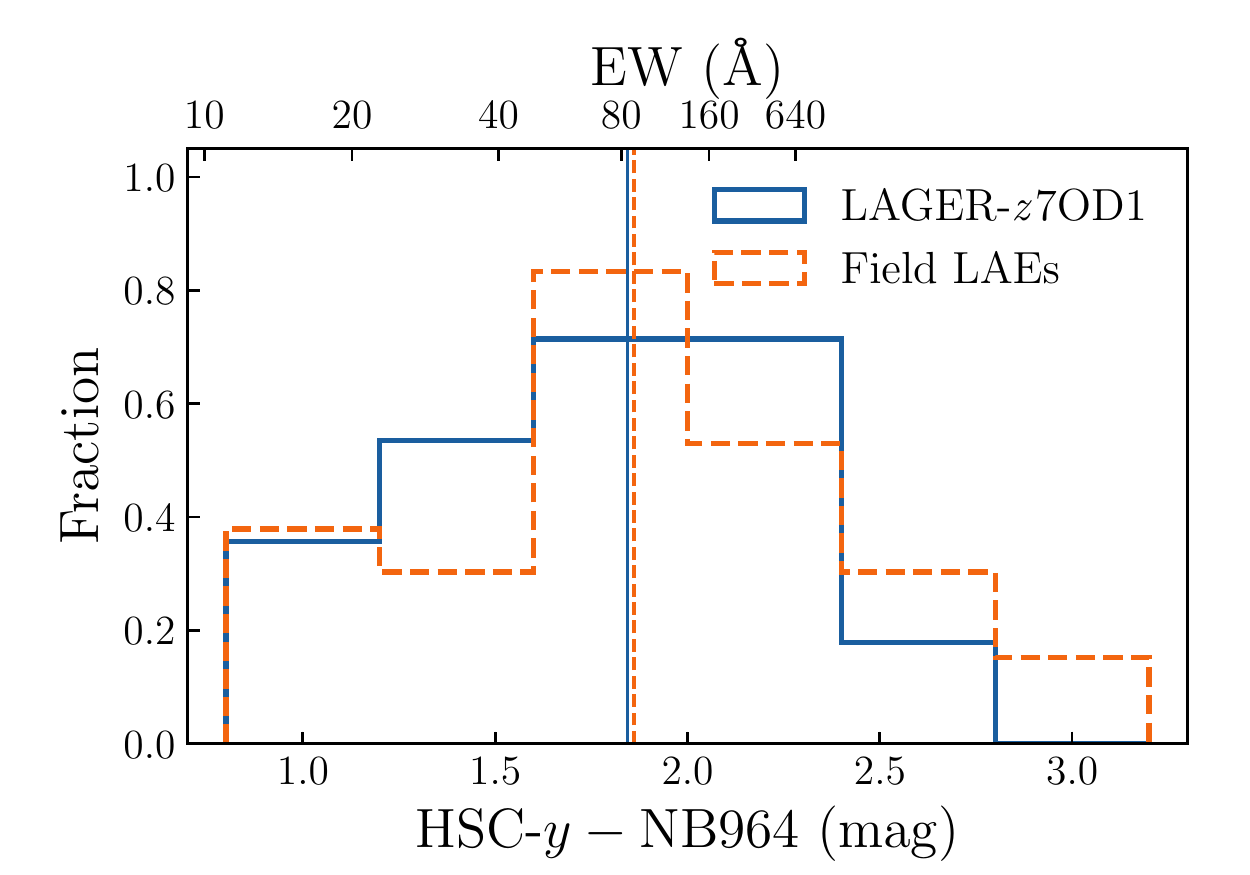}
  \caption{{\textbf{The HSC-$y$ -- DECam-NB964 (and \lya\ EW) distribution of LAEs in the LAGER COSMOS field.}
  The LAEs inside the \pc\ are plotted in blue and those field LAEs in orange. For sources without detection in HSC-$y$ we simply adopt the 2 $\sigma$ lower limits to their HSC-$y$ magnitudes. Most sources with color $>$ 2 in the plot are non-detected in HSC-$y$. The vertical lines plot the median colors (1.84 and 1.86) respectively.
  The tick mark of \lya\ EW is derived from the color assuming a redshift of 6.931 (corresponding to the center of NB964 transmission curve).}
  \label{fig4}
  }
\end{figure}

  \subsection*{EW/color distribution:}
  The \lya\ EWs of narrowband selected LAEs can be well represented by the color between narrowband and the underlying broadband. 
  For LAEs with spectroscopic redshifts one could derive more precise \lya\ EW measurements, after correcting for the wavelength dependence (non-boxcar shape) of the narrowband transmission and the redshift dependence of continuum contribution to narrowband photometry (see Fig. 2 of H19).
  As not all LAEs (particularly those field LAEs) have spectroscopic redshifts, here, we simply use the HSC-$y$ -- DECam-NB964 color as an indicator of \lya\ EW and compare the color distribution of LAEs inside the \pc\ with those field LAEs (Extended Data Fig. 1). Kolmogorov–Smirnov test shows no statistical difference between two samples.

\end{methods}


\begin{thebibliography}{10}
  \expandafter\ifx\csname url\endcsname\relax
    \def\url#1{\texttt{#1}}\fi
  \expandafter\ifx\csname urlprefix\endcsname\relax\def\urlprefix{URL }\fi
  \providecommand{\bibinfo}[2]{#2}
  
  \bibitem{Ouchi2005}
\bibinfo{author}{{Ouchi}, M.} \emph{et~al.}
\newblock \bibinfo{title}{{The Discovery of Primeval Large-Scale Structures
  with Forming Clusters at Redshift 6}}.
\newblock \emph{\bibinfo{journal}{\apjl}} \textbf{\bibinfo{volume}{620}},
  \bibinfo{pages}{L1--L4} (\bibinfo{year}{2005}).

\bibitem{Wang2005}
\bibinfo{author}{{Wang}, J.~X.}, \bibinfo{author}{{Malhotra}, S.} \&
  \bibinfo{author}{{Rhoads}, J.~E.}
\newblock \bibinfo{title}{{An Overdensity of Ly{$\alpha$} Emitters at Redshift
  z{\ensuremath{\sim}}5.7 near the Hubble Ultra Deep Field}}.
\newblock \emph{\bibinfo{journal}{\apjl}} \textbf{\bibinfo{volume}{622}},
  \bibinfo{pages}{L77--L80} (\bibinfo{year}{2005}).

\bibitem{Malhotra2005}
\bibinfo{author}{Malhotra, S.} \emph{et~al.}
\newblock \bibinfo{title}{An overdensity of galaxies at z= 5.9 {\ensuremath{\pm}} 0.2
  in the hubble ultra deep field confirmed using the {ACS} grism}.
\newblock \emph{\bibinfo{journal}{\apj}}
  \textbf{\bibinfo{volume}{626}}, \bibinfo{pages}{666--679}
  (\bibinfo{year}{2005}).

\bibitem{Jiang2018}
\bibinfo{author}{{Jiang}, L.} \emph{et~al.}
\newblock \bibinfo{title}{{A giant protocluster of galaxies at redshift 5.7}}.
\newblock \emph{\bibinfo{journal}{Nat. Astron.}}
  \textbf{\bibinfo{volume}{2}}, \bibinfo{pages}{962--966}
  (\bibinfo{year}{2018}).

\bibitem{Harikane2019}
\bibinfo{author}{{Harikane}, Y.} \emph{et~al.}
\newblock \bibinfo{title}{{SILVERRUSH. VIII. Spectroscopic Identifications of
  Early Large-scale Structures with Protoclusters over 200 Mpc at
  z {\ensuremath{\sim}} 6--7: Strong Associations of Dusty Star-forming
  Galaxies}}.
\newblock \emph{\bibinfo{journal}{\apj}} \textbf{\bibinfo{volume}{883}},
  \bibinfo{pages}{142} (\bibinfo{year}{2019}).

\bibitem{Calvi2019}
\bibinfo{author}{{Calvi}, R.} \emph{et~al.}
\newblock \bibinfo{title}{{MOS spectroscopy of protocluster candidate galaxies
  at z = 6.5}}.
\newblock \emph{\bibinfo{journal}{\mnras}} \textbf{\bibinfo{volume}{489}},
  \bibinfo{pages}{3294--3306} (\bibinfo{year}{2019}).

\bibitem{Robertson2015}
\bibinfo{author}{{Robertson}, B.~E.}, \bibinfo{author}{{Ellis}, R.~S.},
  \bibinfo{author}{{Furlanetto}, S.~R.} \& \bibinfo{author}{{Dunlop}, J.~S.}
\newblock \bibinfo{title}{{Cosmic Reionization and Early Star-forming Galaxies:
  A Joint Analysis of New Constraints from Planck and the Hubble Space
  Telescope}}.
\newblock \emph{\bibinfo{journal}{\apjl}} \textbf{\bibinfo{volume}{802}},
  \bibinfo{pages}{L19} (\bibinfo{year}{2015}).

\bibitem{Kulkarni2019}
\bibinfo{author}{{Kulkarni}, G.} \emph{et~al.}
\newblock \bibinfo{title}{{Large Ly {\ensuremath{\alpha}} opacity fluctuations
  and low CMB {\ensuremath{\tau}} in models of late reionization with large
  islands of neutral hydrogen extending to z $<$ 5.5}}.
\newblock \emph{\bibinfo{journal}{\mnras}} \textbf{\bibinfo{volume}{485}},
  \bibinfo{pages}{L24--L28} (\bibinfo{year}{2019}).

\bibitem{Castellano2018}
\bibinfo{author}{{Castellano}, M.} \emph{et~al.}
\newblock \bibinfo{title}{{Spectroscopic Investigation of a Reionized Galaxy
  Overdensity at z = 7}}.
\newblock \emph{\bibinfo{journal}{\apjl}} \textbf{\bibinfo{volume}{863}},
  \bibinfo{pages}{L3} (\bibinfo{year}{2018}).

\bibitem{Tilvi2020}
\bibinfo{author}{{Tilvi}, V.} \emph{et~al.}
\newblock \bibinfo{title}{{Onset of Cosmic Reionization: Evidence of an Ionized
  Bubble Merely 680 Myr after the Big Bang}}.
\newblock \emph{\bibinfo{journal}{\apjl}} \textbf{\bibinfo{volume}{891}},
  \bibinfo{pages}{L10} (\bibinfo{year}{2020}).

\bibitem{Zheng2017}
\bibinfo{author}{{Zheng}, Z.-Y.} \emph{et~al.}
\newblock \bibinfo{title}{{First Results from the Lyman Alpha Galaxies in the
  Epoch of Reionization (LAGER) Survey: Cosmological Reionization at z {\ensuremath{\sim}}
  7}}.
\newblock \emph{\bibinfo{journal}{\apjl}} \textbf{\bibinfo{volume}{842}},
  \bibinfo{pages}{L22} (\bibinfo{year}{2017}).

\bibitem{Itoh2018}
\bibinfo{author}{{Itoh}, R.} \emph{et~al.}
\newblock \bibinfo{title}{{CHORUS. II. Subaru/HSC Determination of the
  Ly{\ensuremath{\alpha}} Luminosity Function at z = 7.0: Constraints on Cosmic
  Reionization Model Parameter}}.
\newblock \emph{\bibinfo{journal}{\apj}} \textbf{\bibinfo{volume}{867}},
  \bibinfo{pages}{46} (\bibinfo{year}{2018}).

\bibitem{Hu2019}
\bibinfo{author}{{Hu}, W.} \emph{et~al.}
\newblock \bibinfo{title}{{The Ly{\ensuremath{\alpha}} Luminosity Function and
  Cosmic Reionization at z {\ensuremath{\sim}} 7.0: A Tale of Two LAGER
  Fields}}.
\newblock \emph{\bibinfo{journal}{\apj}} \textbf{\bibinfo{volume}{886}},
  \bibinfo{pages}{90} (\bibinfo{year}{2019}).

\bibitem{Hu2017}
\bibinfo{author}{{Hu}, W.} \emph{et~al.}
\newblock \bibinfo{title}{{First Spectroscopic Confirmations of z {\ensuremath{\sim}} 7.0
  Ly{$\alpha$} Emitting Galaxies in the LAGER Survey}}.
\newblock \emph{\bibinfo{journal}{\apjl}} \textbf{\bibinfo{volume}{845}},
  \bibinfo{pages}{L16} (\bibinfo{year}{2017}).

\bibitem{Overzier2009}
\bibinfo{author}{{Overzier}, R.~A.} \emph{et~al.}
\newblock \bibinfo{title}{{{\ensuremath{\Lambda}}CDM predictions for galaxy
  protoclusters - I. The relation between galaxies, protoclusters and quasars
  at z \ensuremath{\sim} 6}}.
\newblock \emph{\bibinfo{journal}{\mnras}} \textbf{\bibinfo{volume}{394}},
  \bibinfo{pages}{577--594} (\bibinfo{year}{2009}).

\bibitem{Chiang2013}
\bibinfo{author}{{Chiang}, Y.-K.}, \bibinfo{author}{{Overzier}, R.} \&
  \bibinfo{author}{{Gebhardt}, K.}
\newblock \bibinfo{title}{{Ancient Light from Young Cosmic Cities: Physical and
  Observational Signatures of Galaxy Proto-clusters}}.
\newblock \emph{\bibinfo{journal}{\apj}} \textbf{\bibinfo{volume}{779}},
  \bibinfo{pages}{127} (\bibinfo{year}{2013}).

\bibitem{Merritt1987}
\bibinfo{author}{{Merritt}, D.}
\newblock \bibinfo{title}{{The Distribution of Dark Matter in the Coma
  Cluster}}.
\newblock \emph{\bibinfo{journal}{\apj}} \textbf{\bibinfo{volume}{313}},
  \bibinfo{pages}{121} (\bibinfo{year}{1987}).

\bibitem{Cai2017}
\bibinfo{author}{{Cai}, Z.} \emph{et~al.}
\newblock \bibinfo{title}{{Mapping the Most Massive Overdensities through
  Hydrogen (MAMMOTH). II. Discovery of the Extremely Massive Overdensity
  BOSS1441 at z = 2.32}}.
\newblock \emph{\bibinfo{journal}{\apj}} \textbf{\bibinfo{volume}{839}},
  \bibinfo{pages}{131} (\bibinfo{year}{2017}).

\bibitem{Chanchaiworawit2019}
\bibinfo{author}{{Chanchaiworawit}, K.} \emph{et~al.}
\newblock \bibinfo{title}{{Physical Properties of a Coma-analog Protocluster at
  z = 6.5}}.
\newblock \emph{\bibinfo{journal}{\apj}} \textbf{\bibinfo{volume}{877}},
  \bibinfo{pages}{51} (\bibinfo{year}{2019}).

\bibitem{Mo1996}
\bibinfo{author}{{Mo}, H.~J.} \& \bibinfo{author}{{White}, S.~D.~M.}
\newblock \bibinfo{title}{{An analytic model for the spatial clustering of dark
  matter haloes}}.
\newblock \emph{\bibinfo{journal}{\mnras}} \textbf{\bibinfo{volume}{282}},
  \bibinfo{pages}{347--361} (\bibinfo{year}{1996}).

\bibitem{Jenkins2001}
\bibinfo{author}{{Jenkins}, A.} \emph{et~al.}
\newblock \bibinfo{title}{{The mass function of dark matter haloes}}.
\newblock \emph{\bibinfo{journal}{\mnras}} \textbf{\bibinfo{volume}{321}},
  \bibinfo{pages}{372--384} (\bibinfo{year}{2001}).

\bibitem{Gnedin2000}
\bibinfo{author}{{Gnedin}, N.~Y.}
\newblock \bibinfo{title}{{Cosmological Reionization by Stellar Sources}}.
\newblock \emph{\bibinfo{journal}{\apj}} \textbf{\bibinfo{volume}{535}},
  \bibinfo{pages}{530--554} (\bibinfo{year}{2000}).

\bibitem{RodriguezEspinosa2020}
\bibinfo{author}{{Rodr{\'\i}guez Espinosa}, J.~M.} \emph{et~al.}
\newblock \bibinfo{title}{{An ionized superbubble powered by a protocluster at
  z = 6.5}}.
\newblock \emph{\bibinfo{journal}{\mnras}} \textbf{\bibinfo{volume}{495}},
  \bibinfo{pages}{L17--L21} (\bibinfo{year}{2020}).

\bibitem{Malhotra2006}
\bibinfo{author}{{Malhotra}, S.} \& \bibinfo{author}{{Rhoads}, J.~E.}
\newblock \bibinfo{title}{{The Volume Fraction of Ionized Intergalactic Gas at
  Redshift z=6.5}}.
\newblock \emph{\bibinfo{journal}{\apjl}} \textbf{\bibinfo{volume}{647}},
  \bibinfo{pages}{L95--L98} (\bibinfo{year}{2006}).

\bibitem{Dijkstra2007b}
\bibinfo{author}{{Dijkstra}, M.}, \bibinfo{author}{{Lidz}, A.} \&
  \bibinfo{author}{{Wyithe}, J. S.~B.}
\newblock \bibinfo{title}{{The impact of The IGM on high-redshift
  Ly{\ensuremath{\alpha}} emission lines}}.
\newblock \emph{\bibinfo{journal}{\mnras}} \textbf{\bibinfo{volume}{377}},
  \bibinfo{pages}{1175--1186} (\bibinfo{year}{2007}).

\bibitem{Yajima2018}
\bibinfo{author}{{Yajima}, H.}, \bibinfo{author}{{Sugimura}, K.} \&
  \bibinfo{author}{{Hasegawa}, K.}
\newblock \bibinfo{title}{{Modelling of Lyman-alpha emitting galaxies and
  ionized bubbles at the epoch of reionization}}.
\newblock \emph{\bibinfo{journal}{\mnras}} \textbf{\bibinfo{volume}{477}},
  \bibinfo{pages}{5406--5421} (\bibinfo{year}{2018}).

\bibitem{Wyithe2015}
\bibinfo{author}{{Wyithe}, S.}, \bibinfo{author}{{Geil}, P.} \&
  \bibinfo{author}{{Kim}, H.}
\newblock \bibinfo{title}{{Imaging HII Regions from Galaxies and Quasars During
  Reionisation with SKA}}.
\newblock \emph{\bibinfo{journal}{Proc. Sci.}} \textbf{\bibinfo{volume}{AASKA14}},
\bibinfo{pages}{015} (\bibinfo{year}{2015}).

\bibitem{Yajima2015}
\bibinfo{author}{{Yajima}, H.}, \bibinfo{author}{{Shlosman}, I.},
  \bibinfo{author}{{Romano-D{\'\i}az}, E.} \& \bibinfo{author}{{Nagamine}, K.}
\newblock \bibinfo{title}{{Observational properties of simulated galaxies in
  overdense and average regions at redshifts z \ensuremath{\simeq} 6--12}}.
\newblock \emph{\bibinfo{journal}{\mnras}} \textbf{\bibinfo{volume}{451}},
  \bibinfo{pages}{418--432} (\bibinfo{year}{2015}).

\bibitem{Lee2017}
\bibinfo{author}{{Lee}, C.~T.} \emph{et~al.}
\newblock \bibinfo{title}{{Properties of dark matter haloes as a function of
  local environment density}}.
\newblock \emph{\bibinfo{journal}{\mnras}} \textbf{\bibinfo{volume}{466}},
  \bibinfo{pages}{3834--3858} (\bibinfo{year}{2017}).

\bibitem{Maio2016}
\bibinfo{author}{{Maio}, U.}, \bibinfo{author}{{Petkova}, M.},
  \bibinfo{author}{{De Lucia}, G.} \& \bibinfo{author}{{Borgani}, S.}
\newblock \bibinfo{title}{{Radiative feedback and cosmic molecular gas: the
  role of different radiative sources}}.
\newblock \emph{\bibinfo{journal}{\mnras}} \textbf{\bibinfo{volume}{460}},
  \bibinfo{pages}{3733--3752} (\bibinfo{year}{2016}).
  
  \setcounter{firstbib}{\value{enumiv}}
  \end{thebibliography}

\begin{thebibliography}{10}
  \setcounter{enumiv}{\value{firstbib}} 
  \expandafter\ifx\csname url\endcsname\relax
    \def\url#1{\texttt{#1}}\fi
  \expandafter\ifx\csname urlprefix\endcsname\relax\def\urlprefix{URL }\fi
  \providecommand{\bibinfo}[2]{#2}

  \bibitem{Planck2018VI}
\bibinfo{author}{{Planck Collaboration}} \emph{et~al.}
\newblock \bibinfo{title}{{Planck 2018 results. VI. Cosmological parameters}}.
\newblock \emph{\bibinfo{journal}{\aap}} \textbf{\bibinfo{volume}{641}},
  \bibinfo{pages}{A6} (\bibinfo{year}{2020}).

\bibitem{Malhotra2004}
\bibinfo{author}{{Malhotra}, S.} \& \bibinfo{author}{{Rhoads}, J.~E.}
\newblock \bibinfo{title}{{Luminosity Functions of Ly{$\alpha$} Emitters at
  Redshifts z=6.5 and z=5.7: Evidence against Reionization at z$\leq$6.5}}.
\newblock \emph{\bibinfo{journal}{\apjl}} \textbf{\bibinfo{volume}{617}},
  \bibinfo{pages}{L5--L8} (\bibinfo{year}{2004}).

\bibitem{Furlanetto2006A}
\bibinfo{author}{{Furlanetto}, S.~R.}, \bibinfo{author}{{Zaldarriaga}, M.} \&
  \bibinfo{author}{{Hernquist}, L.}
\newblock \bibinfo{title}{{The effects of reionization on Ly{$\alpha$} galaxy
  surveys}}.
\newblock \emph{\bibinfo{journal}{\mnras}} \textbf{\bibinfo{volume}{365}},
  \bibinfo{pages}{1012--1020} (\bibinfo{year}{2006}).

\bibitem{McQuinn2007}
\bibinfo{author}{{McQuinn}, M.}, \bibinfo{author}{{Hernquist}, L.},
  \bibinfo{author}{{Zaldarriaga}, M.} \& \bibinfo{author}{{Dutta}, S.}
\newblock \bibinfo{title}{{Studying reionization with Ly{$\alpha$} emitters}}.
\newblock \emph{\bibinfo{journal}{\mnras}} \textbf{\bibinfo{volume}{381}},
  \bibinfo{pages}{75--96} (\bibinfo{year}{2007}).

\bibitem{Ouchi2010}
\bibinfo{author}{{Ouchi}, M.} \emph{et~al.}
\newblock \bibinfo{title}{{Statistics of 207 Ly{$\alpha$} Emitters at a
  Redshift Near 7: Constraints on Reionization and Galaxy Formation Models}}.
\newblock \emph{\bibinfo{journal}{\apj}} \textbf{\bibinfo{volume}{723}},
  \bibinfo{pages}{869--894} (\bibinfo{year}{2010}).

\bibitem{Hu2010}
\bibinfo{author}{{Hu}, E.~M.} \emph{et~al.}
\newblock \bibinfo{title}{{An Atlas of z = 5.7 and z = 6.5 Ly{$\alpha$}
  Emitters}}.
\newblock \emph{\bibinfo{journal}{\apj}} \textbf{\bibinfo{volume}{725}},
  \bibinfo{pages}{394--423} (\bibinfo{year}{2010}).

\bibitem{Jiang2017}
\bibinfo{author}{{Jiang}, L.} \emph{et~al.}
\newblock \bibinfo{title}{{A Magellan M2FS Spectroscopic Survey of Galaxies at
  5.5 {\ensuremath{<}} z {\ensuremath{<}} 6.8: Program Overview and a Sample of the Brightest
  Ly{$\alpha$} Emitters}}.
\newblock \emph{\bibinfo{journal}{\apj}} \textbf{\bibinfo{volume}{846}},
  \bibinfo{pages}{134} (\bibinfo{year}{2017}).

\bibitem{Chanchaiworawit2017}
\bibinfo{author}{{Chanchaiworawit}, K.} \emph{et~al.}
\newblock \bibinfo{title}{{Gran Telescopio Canarias observations of an
  overdense region of Lyman {\ensuremath{\alpha}} emitters at z = 6.5}}.
\newblock \emph{\bibinfo{journal}{\mnras}} \textbf{\bibinfo{volume}{469}},
  \bibinfo{pages}{2646--2661} (\bibinfo{year}{2017}).

\bibitem{Konno2018}
\bibinfo{author}{{Konno}, A.} \emph{et~al.}
\newblock \bibinfo{title}{{SILVERRUSH. IV. Ly{$\alpha$} luminosity functions at
  z = 5.7 and 6.6 studied with {\ensuremath{\sim}}1300 Ly{$\alpha$} emitters on the 14-21
  deg$^{2}$ sky}}.
\newblock \emph{\bibinfo{journal}{\pasj}} \textbf{\bibinfo{volume}{70}},
  \bibinfo{pages}{S16} (\bibinfo{year}{2018}).

\bibitem{Higuchi2019}
\bibinfo{author}{{Higuchi}, R.} \emph{et~al.}
\newblock \bibinfo{title}{{SILVERRUSH. VII. Subaru/HSC Identifications of
  Protocluster Candidates at z {\ensuremath{\sim}} 6--7: Implications for
  Cosmic Reionization}}.
\newblock \emph{\bibinfo{journal}{\apj}} \textbf{\bibinfo{volume}{879}},
  \bibinfo{pages}{28} (\bibinfo{year}{2019}).

\bibitem{Taylor2020}
\bibinfo{author}{{Taylor}, A.~J.}, \bibinfo{author}{{Barger}, A.~J.},
  \bibinfo{author}{{Cowie}, L.~L.}, \bibinfo{author}{{Hu}, E.~M.} \&
  \bibinfo{author}{{Songaila}, A.}
\newblock \bibinfo{title}{{The Ultraluminous Ly{\ensuremath{\alpha}} Luminosity
  Function at z = 6.6}}.
\newblock \emph{\bibinfo{journal}{\apj}} \textbf{\bibinfo{volume}{895}},
  \bibinfo{pages}{132} (\bibinfo{year}{2020}).

\bibitem{Jung2020}
\bibinfo{author}{{Jung}, I.} \emph{et~al.}
\newblock \bibinfo{title}{{Texas Spectroscopic Search for Ly$\alpha$ Emission
  at the End of Reionization III. the Ly$\alpha$ Equivalent-width Distribution
  and Ionized Structures at $z > 7$}}.
\newblock \emph{\bibinfo{journal}{arXiv e-prints}}
  \bibinfo{pages}{arXiv:2009.10092} (\bibinfo{year}{2020}).

\bibitem{Zheng2019}
\bibinfo{author}{{Zheng}, Z.-Y.} \emph{et~al.}
\newblock \bibinfo{title}{{Design for the First Narrowband Filter for the Dark
  Energy Camera: Optimizing the LAGER Survey for z {\ensuremath{\sim}} 7
  Galaxies}}.
\newblock \emph{\bibinfo{journal}{\pasp}} \textbf{\bibinfo{volume}{131}},
  \bibinfo{pages}{074502} (\bibinfo{year}{2019}).

\bibitem{Aihara2018}
\bibinfo{author}{{Aihara}, H.} \emph{et~al.}
\newblock \bibinfo{title}{{First data release of the Hyper Suprime-Cam Subaru
  Strategic Program}}.
\newblock \emph{\bibinfo{journal}{\pasj}} \textbf{\bibinfo{volume}{70}},
  \bibinfo{pages}{S8} (\bibinfo{year}{2018}).

\bibitem{Dressler2011}
\bibinfo{author}{{Dressler}, A.} \emph{et~al.}
\newblock \bibinfo{title}{{IMACS: The Inamori-Magellan Areal Camera and
  Spectrograph on Magellan-Baade}}.
\newblock \emph{\bibinfo{journal}{\pasp}} \textbf{\bibinfo{volume}{123}},
  \bibinfo{pages}{288} (\bibinfo{year}{2011}).

\bibitem{Oemler2017}
\bibinfo{author}{{Oemler}, A.}, \bibinfo{author}{{Clardy}, K.},
  \bibinfo{author}{{Kelson}, D.}, \bibinfo{author}{{Walth}, G.} \&
  \bibinfo{author}{{Villanueva}, E.}
\newblock \bibinfo{title}{{COSMOS: Carnegie Observatories System for
  MultiObject Spectroscopy}} (\bibinfo{year}{2017}).

\bibitem{Kashikawa2006}
\bibinfo{author}{{Kashikawa}, N.} \emph{et~al.}
\newblock \bibinfo{title}{{The End of the Reionization Epoch Probed by
  Ly{\ensuremath{\alpha}} Emitters at z = 6.5 in the Subaru Deep Field}}.
\newblock \emph{\bibinfo{journal}{\apj}} \textbf{\bibinfo{volume}{648}},
  \bibinfo{pages}{7--22} (\bibinfo{year}{2006}).

\bibitem{Ning2020}
\bibinfo{author}{{Ning}, Y.} \emph{et~al.}
\newblock \bibinfo{title}{{The Magellan M2FS Spectroscopic Survey of
  High-redshift Galaxies: A Sample of 260 Ly{\ensuremath{\alpha}} Emitters at
  Redshift z {\ensuremath{\approx}} 5.7}}.
\newblock \emph{\bibinfo{journal}{\apj}} \textbf{\bibinfo{volume}{903}},
  \bibinfo{pages}{4} (\bibinfo{year}{2020}).

\bibitem{Steidel1998}
\bibinfo{author}{{Steidel}, C.~C.} \emph{et~al.}
\newblock \bibinfo{title}{{A Large Structure of Galaxies at Redshift Z
  approximately 3 and Its Cosmological Implications}}.
\newblock \emph{\bibinfo{journal}{\apj}} \textbf{\bibinfo{volume}{492}},
  \bibinfo{pages}{428--438} (\bibinfo{year}{1998}).

\bibitem{Ouchi2018}
\bibinfo{author}{{Ouchi}, M.} \emph{et~al.}
\newblock \bibinfo{title}{{Systematic Identification of LAEs for Visible
  Exploration and Reionization Research Using Subaru HSC (SILVERRUSH). I.
  Program strategy and clustering properties of {\ensuremath{\sim}}2000 Ly{$\alpha$}
  emitters at z = 6--7 over the 0.3--0.5 Gpc$^{2}$ survey area}}.
\newblock \emph{\bibinfo{journal}{\pasj}} \textbf{\bibinfo{volume}{70}},
  \bibinfo{pages}{S13} (\bibinfo{year}{2018}).

\bibitem{Leitherer1999}
\bibinfo{author}{{Leitherer}, C.} \emph{et~al.}
\newblock \bibinfo{title}{{Starburst99: Synthesis Models for Galaxies with
  Active Star Formation}}.
\newblock \emph{\bibinfo{journal}{\apjs}} \textbf{\bibinfo{volume}{123}},
  \bibinfo{pages}{3--40} (\bibinfo{year}{1999}).

\bibitem{Cen2000}
\bibinfo{author}{{Cen}, R.} \& \bibinfo{author}{{Haiman}, Z.}
\newblock \bibinfo{title}{{Quasar Str{\"o}mgren Spheres Before Cosmological
  Reionization}}.
\newblock \emph{\bibinfo{journal}{\apjl}} \textbf{\bibinfo{volume}{542}},
  \bibinfo{pages}{L75--L78} (\bibinfo{year}{2000}).

\bibitem{Vanzella2010}
\bibinfo{author}{{Vanzella}, E.} \emph{et~al.}
\newblock \bibinfo{title}{{The Great Observatories Origins Deep Survey:
  Constraints on the Lyman Continuum Escape Fraction Distribution of
  Lyman-break Galaxies at 3.4 $<$ z $<$ 4.5}}.
\newblock \emph{\bibinfo{journal}{\apj}} \textbf{\bibinfo{volume}{725}},
  \bibinfo{pages}{1011--1031} (\bibinfo{year}{2010}).

\bibitem{Grazian2017}
\bibinfo{author}{{Grazian}, A.} \emph{et~al.}
\newblock \bibinfo{title}{{Lyman continuum escape fraction of faint galaxies at
  z {\ensuremath{\sim}} 3.3 in the CANDELS/GOODS-North, EGS, and COSMOS fields with LBC}}.
\newblock \emph{\bibinfo{journal}{\aap}} \textbf{\bibinfo{volume}{602}},
  \bibinfo{pages}{A18} (\bibinfo{year}{2017}).

\bibitem{Bian2020}
\bibinfo{author}{{Bian}, F.} \& \bibinfo{author}{{Fan}, X.}
\newblock \bibinfo{title}{{Lyman continuum escape fraction in Ly
  {\ensuremath{\alpha}} emitters at z {\ensuremath{\simeq}} 3.1}}.
\newblock \emph{\bibinfo{journal}{\mnras}} \textbf{\bibinfo{volume}{493}},
  \bibinfo{pages}{L65--L69} (\bibinfo{year}{2020}).

\bibitem{Finkelstein2019}
\bibinfo{author}{{Finkelstein}, S.~L.} \emph{et~al.}
\newblock \bibinfo{title}{{Conditions for Reionizing the Universe with a Low
  Galaxy Ionizing Photon Escape Fraction}}.
\newblock \emph{\bibinfo{journal}{\apj}} \textbf{\bibinfo{volume}{879}},
  \bibinfo{pages}{36} (\bibinfo{year}{2019}).

\bibitem{Ma2020}
\bibinfo{author}{{Ma}, X.} \emph{et~al.}
\newblock \bibinfo{title}{{No missing photons for reionization: moderate
  ionizing photon escape fractions from the FIRE-2 simulations}}.
\newblock \emph{\bibinfo{journal}{\mnras}}  (\bibinfo{year}{2020}).

\bibitem{Finkelstein2009}
\bibinfo{author}{{Finkelstein}, S.~L.}, \bibinfo{author}{{Rhoads}, J.~E.},
  \bibinfo{author}{{Malhotra}, S.} \& \bibinfo{author}{{Grogin}, N.}
\newblock \bibinfo{title}{{Lyman Alpha Galaxies: Primitive, Dusty, or
  Evolved?}}
\newblock \emph{\bibinfo{journal}{\apj}} \textbf{\bibinfo{volume}{691}},
  \bibinfo{pages}{465--481} (\bibinfo{year}{2009}).

\bibitem{Lai2008}
\bibinfo{author}{{Lai}, K.} \emph{et~al.}
\newblock \bibinfo{title}{{Spitzer Constraints on the Stellar Populations of
  Ly{\ensuremath{\alpha}}-Emitting Galaxies at z = 3.1}}.
\newblock \emph{\bibinfo{journal}{\apj}} \textbf{\bibinfo{volume}{674}},
  \bibinfo{pages}{70--74} (\bibinfo{year}{2008}).

\bibitem{Cai2014}
\bibinfo{author}{{Cai}, Z.-Y.} \emph{et~al.}
\newblock \bibinfo{title}{{A Physical Model for the Evolving Ultraviolet
  Luminosity Function of High Redshift Galaxies and their Contribution to the
  Cosmic Reionization}}.
\newblock \emph{\bibinfo{journal}{\apj}} \textbf{\bibinfo{volume}{785}},
  \bibinfo{pages}{65} (\bibinfo{year}{2014}).

\end{thebibliography}

\section*{Supplementary Information}

\begin{table*}
  \renewcommand{\tablename}{Supplementary Table}
  \caption{\textbf{Properties of LAEs in the \pc.} We list the 21 member LAEs in \pc. Column 1 lists the source IDs of 21 LAEs. Columns 2 and 3 are the coordinates. Column 4 is the \lya\ photometric luminosity. Column 5 lists the redshifts inferred from the line center for spectroscopic confirmations. Columns 6 -- 8 show their {AUTO} magnitude in the narrowbands DECam-NB964, HSC-NB973, and underlying broadband HSC-y {($2\sigma$ upper limits for non-detections)}. Column 9 lists the bubble size inferred using the relation in ref.\cite{Yajima2018}. Column 10 is the source IDs in ref.\cite{Hu2019} (hereafter H19).}
  \label{tbl1}
  \centering
  \begin{tabular}{c c c c c c c c c c}
  \hline 
  \hline
  \tabincell{c}{ID \\ \ }  & \tabincell{c}{RA \\ \ } & \tabincell{c}{DEC \\ \ } &  \tabincell{c}{$\log\ L_{\mathrm{Ly\alpha}}$ \\ (erg s$^{-1}$) } & \tabincell{c}{Redshift \\ \ } & \tabincell{c}{DECam-NB964 \\ (mag)} & \tabincell{c}{HSC-NB973 \\ (mag)} &\tabincell{c}{HSC-$y$ \\ (mag)} & \tabincell{c}{$R$ \\ (cMpc) } & \tabincell{c}{ID in H19\\ \ }\\
  \hline
  \multicolumn{10}{c}{Spectroscopically Confirmed} \\
  \hline
  LAE-1 & 10:02:06.0 & +02:06:46.3 & $43.54^{+0.03}_{-0.03}$ & 6.938 & $23.08\pm0.06$ & 23.77$^c$ & $25.30\pm0.33$ & 14.5 & COSMOS-1\\
  LAE-2 & 10:01:53.5 & +02:04:59.8 & $43.33^{+0.07}_{-0.08}$ & 6.932 & $23.22\pm0.10$ & 24.75$^c$ &$24.11\pm0.12$ & 11.5 & COSMOS-3\\
  LAE-3 & 10:03:10.5 & +02:12:30.8 & $43.49^{+0.04}_{-0.05}$ & 6.923 & $23.17\pm0.08$ & ** &$25.15\pm0.33$ & 13.7 & COSMOS-2\\
  LAE-4 & 10:03:32.7 & +02:09:25.1 & $43.04^{+0.09}_{-0.11}$ & 6.900 & $24.33\pm0.16$ & ** &$26.42\pm0.70$ & 8.5 & COSMOS-10\\
  LAE-5 & 10:03:30.7 & +02:14:08.5 & $43.03^{+0.07}_{-0.08}$ & 6.899 & $24.37\pm0.13$ & ** &$26.63\pm0.58$ & 8.0 & N$^a$\\
  LAE-6 & 10:03:28.0 & +02:08:51.3 & $43.03^{+0.10}_{-0.14}$ & 6.915 & $24.34\pm0.23$ & ** &$26.45\pm0.34$ & 7.3 & N$^a$\\
  LAE-7 & 10:03:05.2 & +02:09:14.7 & $42.79^{+0.12}_{-0.17}$ & 6.945 & $24.69\pm0.19$ & $24.28\pm0.18$ &$25.81\pm0.35$ & 6.1 & N$^b$\\
  LAE-9 & 10:03:16.0 & +02:15:42.3 & $42.70^{+0.13}_{-0.20}$ & 6.920 & $24.95\pm0.21$ & ** &$26.13\pm0.45$ & 6.4 & N$^a$\\
  LAE-10 & 10:02:42.3 & +02:06:55.2 & $42.56^{+0.13}_{-0.18}$ & 6.922 & $25.29\pm0.22$ & ** &$26.42\pm0.28$ & 5.8 & N$^a$\\
  LAE-11 & 10:02:39.4 & +02:07:12.1 & $42.69^{+0.11}_{-0.15}$ & 6.962 & $25.13\pm0.21$ & 24.78$^c$ &$26.84\pm0.53$ & 6.4 & COSMOS-41\\
  LAE-13 & 10:02:33.5 & +02:07:09.5 & $42.68^{+0.12}_{-0.16}$ & 6.936 & $25.14\pm0.20$ & ** &$26.85\pm0.69$ & 6.4 & COSMOS-42\\
  LAE-15 & 10:02:23.4 & +02:05:04.8 & $43.38$$^d$ & 6.971 & $25.04\pm0.19$ & 23.68$^c$ & $26.41\pm0.34$& 12.1 & COSMOS-49\\
  LAE-16 & 10:02:32.9 & +02:05:52.8 & $42.85^{+0.08}_{-0.10}$ & 6.915 & $24.82\pm0.18$ & ** &$>27.2$ & 7.3 &COSMOS-29\\
  LAE-17 & 10:03:33.5 & +02:07:19.8 & $42.94^{+0.09}_{-0.11}$ & 6.917 & $24.61\pm0.20$ & ** &$>27.2$ & 7.9 & COSMOS-17\\
  LAE-18 & 10:03:37.3 & +02:07:36.7 & $42.86^{+0.08}_{-0.10}$ & 6.953 & $24.81\pm0.18$ & ** &$>27.2$ & 7.3 & COSMOS-27\\
  LAE-19 & 10:03:39.3 & +02:07:47.2 & $42.69^{+0.16}_{-0.25}$ & 6.943 & $24.91\pm0.22$ & ** &$25.93\pm0.50$ & 6.4 & COSMOS-34\\
  \hline
  \multicolumn{10}{c}{Not Yet Confirmed} \\  
  \hline
  LAE-8 & 10:02:09.0 & +02:04:11.0 & $42.84^{+0.11}_{-0.14}$ & ** & $24.81\pm0.21$ & ** &$26.8\pm0.62$ & 7.2 & COSMOS-25\\
  LAE-12 & 10:03:00.1 & +02:14:49.5 & $42.81^{+0.10}_{-0.14}$ & ** & $24.76\pm0.20$ & ** &$26.24\pm0.28$ & 7.0 &COSMOS-24\\
  LAE-14 & 10:02:08.3 & +02:06:59.6 & $42.79^{+0.14}_{-0.21}$ & ** & $24.86\pm0.25$ & ** &$26.45\pm0.67$ & 6.9 &COSMOS-30\\
  LAE-20 & 10:02:47.1 & +02:10:40.1 & $43.05$$^d$ & ** & $25.40\pm0.24$ & 24.52$^c$ &$26.86\pm0.44$ & 7.0 &N$^b$\\
  LAE-21 & 10:03:15.6 & +02:18:11.3 & $42.87^{+0.09}_{-0.11}$ & ** & $24.79\pm0.20$ & $25.73\pm0.51$ &$>27.2$ & 6.9 &N$^b$\\
  \hline
  \end{tabular}
  \begin{tablenotes}
     \item[$^a$] $^a$ LAE-5, 6, 9, 10 were not included in H19 as they are labelled as lower-grade candidates for various reasons (LAE-5: close to bad image regions; LAE-6: noisy signal in the NB image; LAE-9: adjacent to a foreground galaxy within $3''$; LAE-10: with DECam-NB964 signal lower than 5$\sigma$), but latterly got spectroscopically confirmed. 
     \item[$^b$] $^b$ LAE-7, 20, and 21 are selected using the stacked image of DECam-NB964 and HSC-NB973 images.
     \item[$^c$]  $^c$ {We adopt the HSC-NB973 magnitudes given in ref.\cite{Itoh2018}.}
     \item[$^d$]  $^d$ {We adopt the \lya\ luminosities given in ref\cite{Itoh2018}, because their \lya\ lines locate in the red tail of the DECam-NB964 and their \lya\ luminosities will be severely underestimated if using DECam-NB964 magnitude. } 
  \end{tablenotes} 
  \end{table*}
  

\addtocounter{figure}{-1} 
\begin{figure*}
\renewcommand{\figurename}{Supplementary Figure}
\centering
\includegraphics[width=7in]{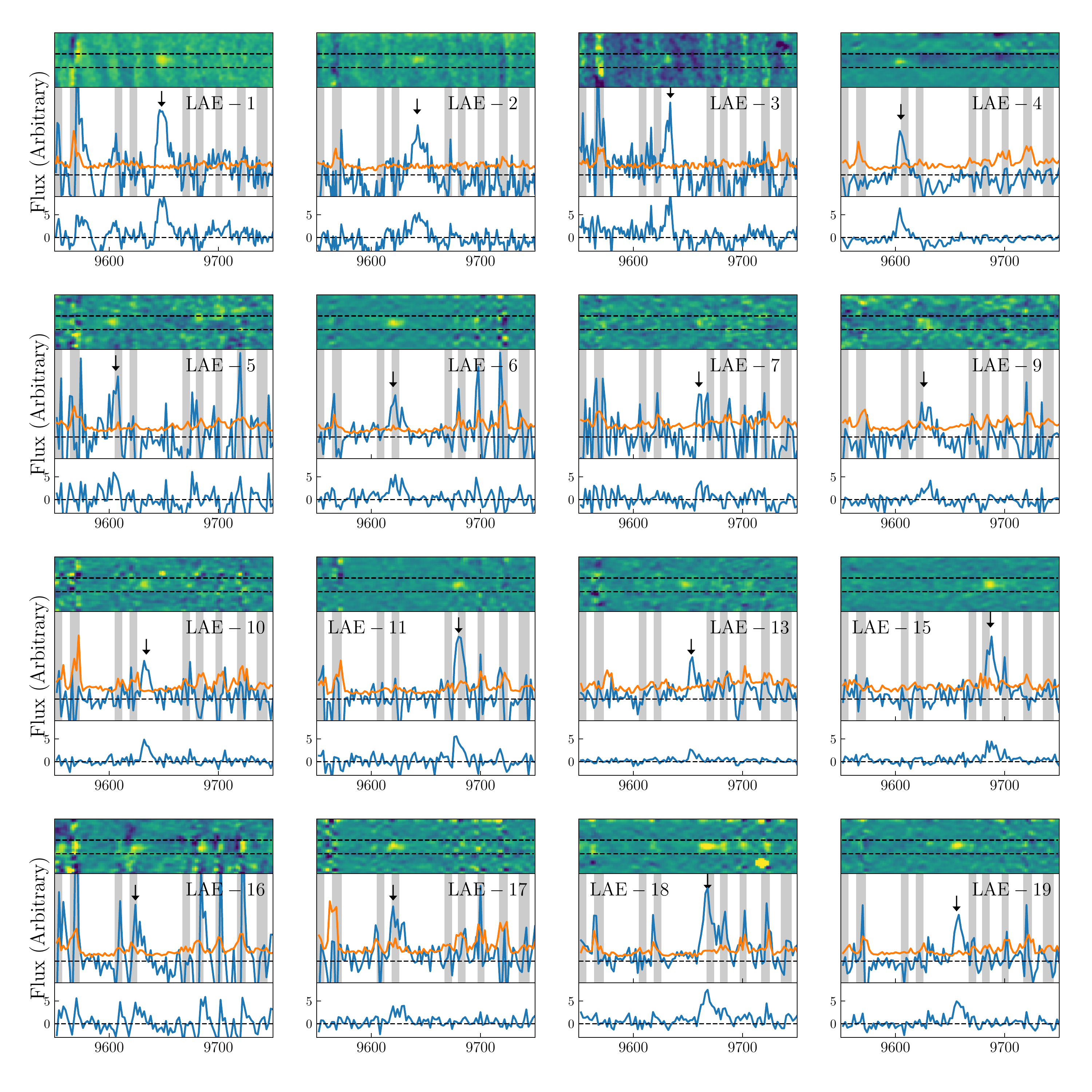}
\caption{\textbf{Two- and one-dimensional spectra of 16 confirmed LAEs in \pc.} 
  In the top panel the two-dimensional spectra (yellow is high flux and blue is low flux) are smoothed by a Gaussian kernel with 1 pixel for better illustration.
  The two black dashed lines (separated by $1''$ vertically) represent the expected slit position of LAEs in the 2D spectra.
  In the middle panel, the blue lines are the one-dimensional spectra and the orange lines are the noise spectra. The grey regions represent the sky OH lines (imperfect sky line subtraction could yield artificial signals visible in the spectra). 
  The dashed horizontal lines indicate zero-flux level and the black arrows mark the peak of the identified \lya\ line profiles.
  In the bottom panel, we plot the S/N spectra with the dashed horizontal lines showing zero S/N.
  Due to the flaws in the slit, LAE-16 shows a noisy 2D spectrum. However, a clear broad red wing of the line is revealed which falls in the skyline free region,  and the line is considerably broader than artificial line signals in the spectrum.
 Thus, we identify it as a \lya\ line.
}
\label{fig4}
\end{figure*}

\end{document}